# Second harmonic generation in nano-structured thin-film lithium niobate waveguides

Cheng Wang,[1,†] Xiao Xiong,[1,2,†] Nicolas Andrade,[1,3] Vivek Venkataraman,[1] Xi-Feng Ren,[2] Guang-Can Guo,[2] and Marko Lončar[1,*]

[1]*John A. Paulson School of Engineering and Applied Sciences, Harvard University, 29 Oxford Street, Cambridge, MA 02138, USA*

[2]*Key Laboratory of Quantum Information & Synergetic Innovation Center of Quantum Information & Quantum Physics, University of Science and Technology of China, Hefei, Anhui 230026, China*

[3]*Department of Electrical and Computer Engineering, Virginia Commonwealth University, Richmond, Virginia 23284, USA*

*[*] loncar@seas.harvard.edu*

[†] These authors contributed equally to this work.

**Abstract:** Integrated thin-film lithium niobate platform has recently emerged as a promising candidate for next-generation, high-efficiency wavelength conversion systems that allow dense packaging and mass-production. Here we demonstrate efficient, phase-matched second harmonic generation in lithographically-defined thin-film lithium niobate waveguides with sub-micron dimensions. Both modal phase matching in fixed-width waveguides and quasi-phase matching in periodically grooved waveguides are theoretically proposed and experimentally demonstrated. Our low-loss (~ 3.0 dB/cm) nanowaveguides possess normalized conversion efficiencies as high as 41% $W^{-1}cm^{-2}$.

## *Introduction*

Second order ($\chi(2)$) nonlinear optical processes, including second harmonic generation (SHG), sum/difference frequency generation (SFG/DFG), and parametric down conversion, not only are crucial for accessing new spectral ranges in classical optics [1-3], but also act as key resources for non-classical light generation in quantum information processing [4-6]. Conventional second order wavelength conversion systems are typically realized in ion-exchanged periodically poled lithium niobate (PPLN) waveguides, where quasi-phase matching is achieved by periodic domain inversion [7-11]. These devices take advantage of lithium niobate's (LiNbO$_3$, LN) large diagonal $\chi^{(2)}$ coefficient ($d_{33}$ = 27 pm/V) and wide transmission window from UV to mid-IR [12]. However, the low index contrast ($\Delta$n ~ 0.02) between waveguide core and cladding usually results in large device dimensions (~ cm long and ~ 10 µm wide) and large bending radii (mm scale) [13], preventing dense integration.

On the other hand, recent years have seen tremendous progress in the field of nonlinear nanophotonics [14-22]. Nonlinear interactions could be enhanced by orders of magnitude due to the superior light confinement in these wavelength-scale devices. Moreover, the use of well-developed nanofabrication methods offers the possibility to build scalable, low-cost and highly integrated nonlinear optical systems.

While LN's linear and nonlinear optical properties offer unique opportunities for novel devices, realization of LN nanophotonics has remained challenging until the recent commercial availability of LN on insulator (LNOI) substrates and development on LN nanofabrication techniques [23]. Since then, various thin-film LN optical components have been realized [24-32]. In particular, micron-scale LN waveguides and their application in nonlinear wavelength conversions have been demonstrated [30-32]. However, these waveguides either suffer from a high propagation loss [30], or have large bending radii due to weak confinement [31, 32].

In this work we demonstrate efficient SHG in low-loss (~ 3.0 dB/cm) thin-film LN nanowaveguides defined by direct dry etching, with normalized conversion efficiencies as high as 41% W$^{-1}$cm$^{-2}$. This is enabled by the ability to precisely engineer the dispersion properties and device dimensions of the LN nanowaveguides by the top-down fabrication method we use. To achieve the phase matching condition in our devices, we theoretically propose and experimentally demonstrate two distinct schemes: (1) modal phase matching between 1$^{st}$ and 3$^{rd}$ order transverse-electric (TE) modes in waveguides of fixed width; (2) quasi-phase matching in periodically grooved lithium niobate (PGLN) waveguides. We show that both methods feature unique advantages and are promising for future integrated nonlinear wavelength conversion systems.

## *Design and simulations*

Figure 1(a) displays the cross-section schematic of a typical *x*-cut thin-film LNOI waveguide cladded in silica. The geometric parameters of such a waveguide include top width $w_t$, bottom width $w_b$, thickness $t$, and sidewall angle $\theta$ (introduced by the dry etching process). The coordinates in Fig. 1(a) are aligned with the crystalline directions of LN, where $z$ is the extraordinary axis. This waveguide geometry supports both TE like and transverse-magnetic (TM) like modes. However, here we are interested in the TE modes only in order to access the largest nonlinear coefficient ($d_{33}$ = 27 pm/V) [12]. Figure 1(b) shows the dependence of effective mode indices ($n_{eff}$) of both fundamental mode at pump wavelength (~ 1550 nm) as well as fundamental and higher order modes at second harmonic (SH) wavelength (~ 775 nm) as a function of waveguide top width ($w_t$). Representative modal profiles of $E_z$ components at both wavelengths are displayed in Fig. 1(c). The results were obtained using a Finite Difference Eigenmode solver (MODE Solutions, Lumerical). In the simulation, we use $t = 400$ nm and $\theta = 40°$, which are taken from actual device dimensions. To achieve phase matching, $n_{eff}$ at both wavelengths need to be equal. In our system this cannot be achieved for fundamental TE modes at pump and SH wavelengths, due to both material and waveguide dispersions. In the following sections, we show two methods to address the phase mismatch issue based on the waveguide dispersion displayed in Fig. 1(b).

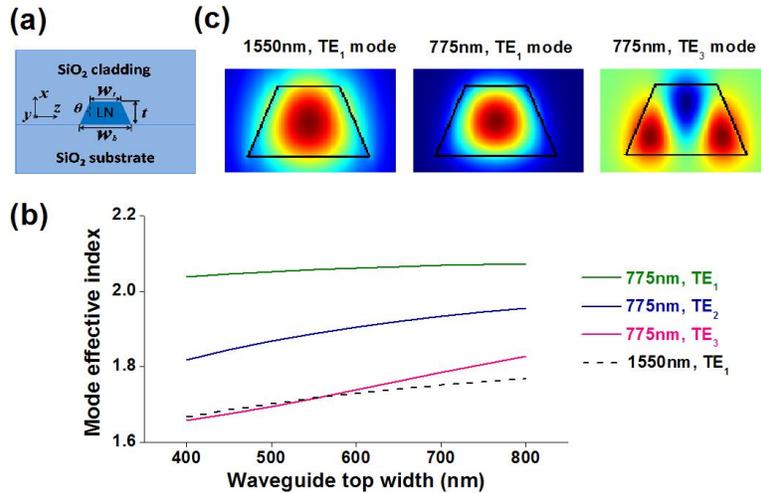

**Figure 1**. (a) Cross-section schematic of the x-cut LNOI waveguide, where the coordinates are aligned with the LN crystal directions. (b) Mode effective indices as a function of waveguide top width at both pump and SH wavelengths. (c) Ez components of the corresponding modes at both wavelengths.

## 2.1 Modal phase matching between 1st and 3rd order TE modes

To achieve modal phase matching, higher order modes at SH wavelength can be used to bring down the $n_{eff}$ to match with that of the fundamental mode at pump wavelength. Note that conversion between $TE_1$ mode at fundamental wavelength and $TE_2$ mode at SH wavelength is prohibited by symmetry. Therefore the lowest possible phase-matching mode at SH wavelength is $TE_3$. In fact, Fig. 1 (b) suggests that phase matching is readily achievable with $w_t \sim 580$ nm.

According to coupled-mode theory and assuming the low-conversion limit [33], the SHG conversion efficiency in a waveguide can be solved as:

$$\gamma = g^2 L^2 P_0 \left(\frac{sin(\delta L/2)}{\delta L/2}\right)^2 \qquad (1)$$

where $\delta = \beta^{2\omega} - 2\beta^{\omega}$ is the phase mismatch, $\beta^q$ is the propagation constant, $q \in \{\omega, 2\omega\}$ represents the corresponding optical frequencies, $L$ is the waveguide length, $P_0$ is the input optical power. Here, the overlap factor is

$$g = \frac{\omega}{4} \iint (E^{2\omega}(x,z))^* d_{33}(x,z)(E^{\omega}(x,z))^2 dxdz \qquad (2)$$

where $E^q(x,z)$ is the normalized electric field distribution in the waveguide cross section, and $d_{33}$ is the diagonal nonlinear coefficient of $LiNbO_3$.

Note that, Eq. (1) has the same form as SHG in a bulk nonlinear crystal. Considering the modal phase matching case ($\delta = 0$), the normalized (by length and input power) conversion efficiency $\eta$ is simply $g^2$.

Using practical waveguide parameters ($w_t$ = 600nm, $w_b$ = 1270 nm, $t$ = 400 nm), and mode profiles for fundamental pump and third order SH mode, we obtain a nonlinear overlap factor $g = 0.774$ $W^{-1/2}cm^{-1}$, which corresponds to a normalized conversion efficiency $\eta = 60\%$ $W^{-1}cm^{-2}$. Taking into account the 3.0 dB/cm waveguide propagation measured in Section 4, a maximum conversion efficiency of 31% $W^{-1}$ can be achieved with a 20 mm long waveguide.

## 2.2 Periodically-grooved lithium niobate (PGLN) waveguides

In conventional PPLN waveguides, quasi-phase matching is realized by periodically inverting the ferroelectric crystal direction [34]. Here, we utilize a periodically-grooved structure to achieve quasi-phase matching between fundamental TE modes at both pump and SH wavelengths. By introducing periodic modulation of the waveguide width, as shown in Fig. 2(a), with a period $\Lambda$, an additional momentum "kick" $\Delta k = 2\pi/\Lambda$ could be applied to the propagating electromagnetic wave, compensating for the phase mismatch $\delta$ [35].

To analytically solve the nonlinear coupled-mode equations in this case, we consider the PGLN waveguide as a perturbation from the uniform waveguide case (described in Section 2.1), with both the linear permittivity $\varepsilon$ and nonlinear coefficient $d_{33}$ periodically modulated along the propagation direction ($y$-axis). This is justified since the effective index perturbations corresponding to 80 nm grove depth are within 2.3% and 0.7% for fundamental and SH wavelengths, respectively. We can therefore expand both $\Delta\varepsilon$ (permittivity difference between waveguide and environment) and $d_{33}$ as Fourier series in $y$:

$$\Delta\varepsilon(x,y,z) = \sum_m \Delta\varepsilon_m(x,z)\exp(-jm\Delta ky) \quad (3)$$

$$d_{33}(x,y,z) = \sum_m d_{33}^{(m)}(x,z)\exp(-jm\Delta ky) \quad (4)$$

In this case, we have the new overlap factor $g'$:

$$g' = g_{NL}^{(1)}(J_0(\varphi_L) + J_2(\varphi_L)) - g_{NL}^{(0)}J_1(\varphi_L) \quad (5)$$

with

$$g_{NL}^{(m)} = \frac{2\omega}{4} \iint (E^{2\omega}(x,z))^* d_{33}^{(m)}(x,z)(E^\omega(x,z))^2 dxdz \quad (6)$$

$$g_L^\omega = \frac{\omega\varepsilon_0}{4} \iint \Delta\varepsilon_1^\omega(x,z)|E^\omega(x,z)|^2 dxdz \quad (7)$$

$$g_L^{2\omega} = \frac{2\omega\varepsilon_0}{4} \iint \Delta\varepsilon_1^{2\omega}(x,z)|E^{2\omega}(x,z)|^2 dxdz \quad (8)$$

$$\varphi_L = 2(g_L^{2\omega} - 2g_L^\omega)/\Delta k \quad (9)$$

Now, the phase mismatch term becomes $\delta' = \beta^{2\omega} - 2\beta^\omega - 2\pi/\Lambda$, and $J_p$ denotes the $p$-th order Bessel function (only first three terms with slow spatial variation are taken). Detailed derivations can be found in Ref. [36].

The new overlap factor $g'$ consists of two terms. The first term results from the periodically varying nonlinear coefficient $d_{33}^{(1)}$, which is the same effect as in PPLN. The second term originates from the constant nonlinear coefficient $d_{33}^{(0)}$ and periodically modulated dielectric constant $\Delta\varepsilon_1$, or grating effect.

For a typical PGLN waveguide ($w_t$ = 670 nm, $w_b$ = 1300 nm, $t$ = 400 nm) with a periodic groove depth of 80 nm, the nonlinear coupling coefficient is calculated to be $g' = 0.345$ W$^{-1/2}$cm$^{-1}$, which corresponds to a normalized conversion efficiency η of 12.1% W$^{-1}$cm$^{-2}$. The total conversion efficiency $\eta$ in the low-conversion limit is plotted as a function of propagation length in Fig. 2(b), in comparison with a uniform waveguide without periodic grooves. Here we also take into consideration the oscillating phase-mismatched optical fields, which could be obtained simply by replacing the overlap factor in Eq. (5) with $G' = g' + g_{NL}^{(0)} J_0(\varphi_L) exp(j\Delta ky)$. The net conversion efficiency features a quadratically increasing envelope with local oscillations.

In a realistic scenario, PGLN waveguide losses should be taken into account. There are two main loss channels in our devices: intrinsic loss due to the leaky nature of the Bloch modes supported by our waveguide, and roughness-induced scattering loss due to fabrication imperfections [37]. We calculated the intrinsic loss component using the Finite Difference Time Domain method (FDTD Solutions, Lumercial) by modeling one unit cell of the corrugated waveguide and applying periodic boundary condition. The results are summarized in Fig. 2(c). As expected, both modes are lossier for deeper grooves, which limits the propagation lengths of the modes and reduces the overall conversion efficiency. Furthermore, the second harmonic mode is a lot more sensitive to the corrugation since the mode has a larger overlap with it (see Fig. 1(c)). On the other hand, larger groove depth ensures stronger nonlinear overlap between the two modes, and can result in larger conversion efficiency (Fig. 2(d)). In order to model this trade-off between the competing effects from the grooves, we introduce loss terms in the nonlinear coupled-mode equations. In addition to the intrinsic loss, we also include additional 3.0 dB/cm fabrication induced scattering loss that we experimentally obtained from our devices, as discussed in following sections.

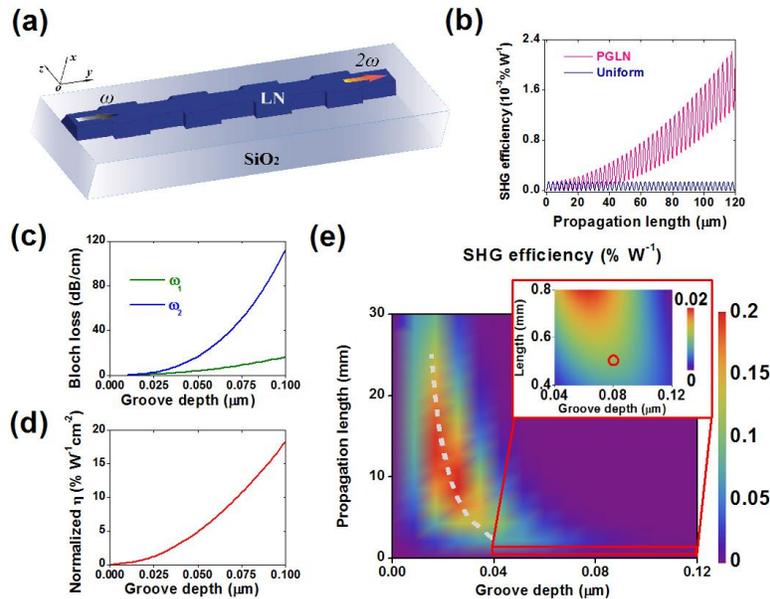

**Figure 2**. (a) 3D cartoon of the proposed PGLN structure. (b) Simulated SHG efficiencies versus propagation length for a PGLN waveguide with a groove depth of 80 nm, in comparison with a uniform LN waveguide. (c) Calculated loss due to the leaky Bloch modes in PGLN at both fundamental and second harmonic wavelengths, as a function of groove depth. (d) Calculated normalized conversion efficiency as a function of groove depth. (e) SHG efficiency dependence on groove depth and propagation length. The global maximum conversion of ~ 0.16% W$^{-1}$ is achieved with ~ 22 nm groove depth and ~ 11mm waveguide length. Inset: Enlarged view for the parameter space in vicinity to our experimental operation point (red circle).The dashed line indicates the optimal groove depth for each propagation length.

Assuming that the pump is not depleted by the generated SH, and that the groove period exactly compensates for the initial phase mismatch, or $\delta' = 0$, the new conversion efficiency including the waveguide loss can be expresses as:

$$\gamma' = g'^2 P_0 \left(\frac{exp(-2\alpha^\omega L) - exp(-\alpha^{2\omega} L)}{2\alpha^\omega - \alpha^{2\omega}}\right)^2 \qquad (10)$$

where $\alpha^\omega, \alpha^{2\omega}$ are the loss coefficients of the PGLN waveguide for pump and SH wavelengths respectively, including both scattering loss and Bloch loss.

Figure 2(e) shows the theoretical result of conversion efficiency as a function of groove depth and waveguide length, with losses taken into account. The global peak in our current device configuration is present for a groove depth ~ 22 nm and a waveguide length ~11 mm, which correspond to a maximum conversion efficiency of 0.16% W$^{-1}$. Note this global maximum efficiency is still limited by the intrinsic propagation loss in our waveguides. For example, if the roughness-induced loss could be reduced to 0.25 dB/cm, a groove depth of 8 nm and a total length of 11 cm could yield a 2.3% W$^{-1}$ overall conversion efficiency. In our proof-of-concept experiments, we focus on shorter waveguides, with a length of 500 µm, which results in an optimal groove depth of ~ 80 nm. In this case, the total waveguide propagation losses are 13.5 dB/cm at pump wavelength and 63.5 dB/cm at SH wavelength, resulting in an overall conversion efficiency of 0.012% W$^{-1}$ and a normalized conversion efficiency of 4.6% W$^{-1}$cm$^{-2}$. Although PGLN waveguide yields lower conversion efficiencies than uniform waveguides, it achieves wavelength conversion between fundamental optical modes at both wavelengths, which is beneficial in many applications.

## *Device fabrication*

Starting from an x-cut LNOI substrate (400 nm thick, NANOLN), uniform and PGLN waveguides were fabricated using a process modified from Ref. [24], except that an amorphous silicon (a-Si) etching mask is used instead of resist. An 800 nm thick a-Si layer was first deposited on the substrate via plasma-enhanced chemical vapor deposition (PECVD). The a-Si layer was then patterned with a combination of electron-beam lithography (EBL) and reactive ion etching (RIE), and used as a hard mask for the subsequent LN dry etching in Ar+ plasma. After removing leftover silicon mask in KOH (80 °C), the waveguides were cladded in silica (3 μm thick) using PECVD. Waveguide facets were diced and polished to ensure high coupling efficiency. Figure 3 shows the scanning electron microscope (SEM) images of uniform LN and PGLN waveguides without cladding. It can be seen that our fabrication process is well capable of delivering designed structures, while maintaining minimum surface roughness and manageable scattering loss.

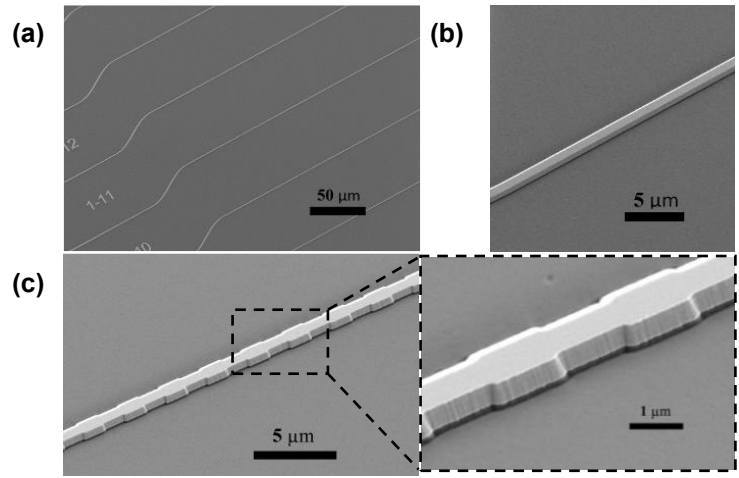

**Figure 3**. Representative scanning electron microscope images of the fabricated devices. (a) An array of LN waveguides with slightly different widths. The bending sections in the waveguides are used to prevent the output fiber from collecting light directly from the input fiber. (b) A typical uniform LN waveguide with fixed width. (c) A typical PGLN waveguide with a spatial modulation period of 2.77 μm and a groove depth of 80 nm.

## *Optical measurements*

We measured the SHG response of both modal phase-matched LN and PGLN waveguides using the setup shown in Fig. 4(a). Tunable telecom lasers (Santec TSL-510, 1480 – 1680 nm) were used as light sources at pump wavelength. A fiber polarization controller was used to ensure TE mode input before end-fire coupling into the device under test. The SHG signals with TM polarized input are usually orders of magnitude lower than TE input due to misaligned crystal relevant axes. Light is coupled into and out of the waveguide facets using tapered lensed fibers. The output light is sent to either an InGaAs photodetector, or a silicon avalanche photodetector (APD), to monitor the linearly transmitted telecom light and the SHG signal respectively.

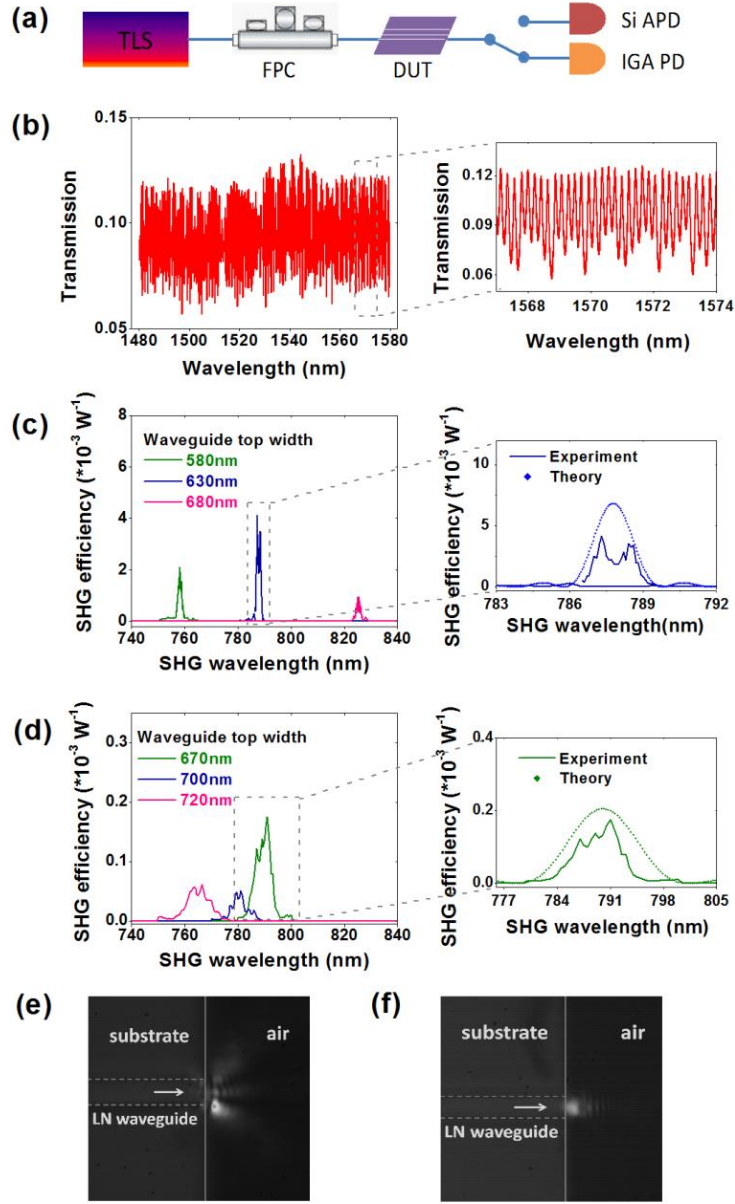

**Figure 4**. (a) Schematic of the measurement setup. Light from the telecom tunable laser source (TLS) is coupled into the device under test (DUT) after passing through a fiber polarization controller (FPC). SHG signal is measured using a silicon avalanche photodetector (Si APD), while linear transmission at telecom wavelength is monitored using an InGaAs photodetector (IGA PD). (b) Transmission spectrum of a typical uniform LN waveguide. Inset: zoom-in view of the wavelength range used for calculating propagation loss. (c-d) Conversion efficiency versus SHG wavelength for (c) uniform LN waveguides and (d) PGLN waveguides with different waveguide widths (measured at the waveguide top). Insets: comparison between experimental (solid) and theoretical (dotted) SHG efficiencies and bandwidths. (e-f) CCD camera images of the scattered SHG light at the output facets of a uniform LN waveguide (e) and a PGLN waveguide (f), indicating the corresponding output optical modes.

Figure 4(b) shows the linear transmission spectrum (in absolute unit) of a typical uniform waveguide at telecom wavelengths. It indicates a total fiber-to-fiber loss ~ 10.2 dB, resulting from a combination of facet coupling loss and waveguide propagation loss. The fringe pattern seen in Fig. 4(b) is a result of Fabry-Perot interference between the two polished facets of the waveguide. From the contrast between maximum and minimum of the transmission fringes, we can extract the waveguide propagation loss using the following equation [38]:

$$\alpha = \frac{4.34}{L}\left(\ln R - \ln \tilde{R}\right) \qquad (11)$$

with $\tilde{R} = \frac{1}{K}\left(1 - \sqrt{1 - K^2}\right)$, $K = \frac{I_{max} - I_{min}}{I_{max} + I_{min}}$. Here $I_{max}$ and $I_{min}$ are the maximum and minimum intensities of the transmitted light. $R$ is the reflection coefficient at the waveguide facet, which is calculated to be 0.147 using FDTD. $L$ is the waveguide length, which is 3 mm in this case (including coupling and bending sections with phase-mismatched widths). The waveguide propagation loss is thus calculated to be 3.0 ± 0.2 dB/cm. This value represents the best propagation loss reported in thin-film LN waveguides with sub-wavelength light confinement ($A_{eff}$ = 0.52 μm$^2$ at telecom wavelength), and is consistent with the quality factor achieved in our previous microdisk resonators [24]. Using the calculated propagation loss and the measured transmission, we are able to estimate the fiber-to-waveguide coupling efficiency. By characterizing many waveguides on the same chip used in the following experiments, an average coupling efficiency of 33% on each facet is extracted (4.8 dB/facet). This coupling loss can be reduced to < 0.5 dB with adiabatic mode transition using tapered fibers [39].

Figures 4(c) and 4(d) show the measured SHG efficiencies as a function of SH wavelength for both modal phase matched LN waveguides (1 mm long) and quasi-phase matched PGLN waveguides (0.5 mm long). These values have accounted for the APD spectral quantum efficiency and the facet coupling loss, and are normalized by the pump laser power. For each scheme, a set of waveguides with slightly different widths are characterized. SHG peaks can be clearly observed at the corresponding (quasi-) phase matching wavelengths. The SHG peak wavelength changes with increasing waveguide width in both cases, but in opposite directions, which agrees with theoretical prediction. For the modal phase matched waveguide with 630 nm top width, we measured SHG power of 1.38 pW at 18.3 μW pump power, which corresponds to a normalized conversion efficiency of 41% W$^{-1}$cm$^{-2}$. Similarly, for PGLN with a top width of 670 nm, the measured output power is 0.334 pW at 44.3 μW pump power, or 6.8% W$^{-1}$cm$^{-2}$. The dotted curves in the insets of Figs. 4(c) and 4(d) show the theoretical SHG bandwidths and peak intensities calculated using the methods described in Section 2. In both schemes, the measured bandwidths match the theoretical predictions well, while the peak efficiencies are slightly lower than theory, indicating possible inhomogeneity in the waveguide dimensions throughout the chip.

To confirm that the generated SH light couple to third-order and fundamental mode in the case of modal and groove-assisted phase matching, respectively, we used a high NA (0.8) lens and a black & white visible CCD camera to monitor the scattered SH light at the output facet [Figs. 4(e) and 4(f)]. In the case of uniform LN waveguide [Fig. 4(e)], the SHG signal clearly radiates in three lobes, indicating SH light being generated in the 3rd order transverse mode. In comparison, the SH light generated in the PGLN waveguide [Fig. 4(f)] is in the fundamental mode and radiates only in one lobe.

## *Conclusions*

To conclude, we have demonstrated efficient nonlinear wavelength conversion in sub-wavelength LN waveguides, which are realized by a top-down fabrication method and can be densely integrated and mass produced. Without corrugation, our waveguides feature a low propagation loss of ~ 3.0 dB/cm, which is crucial for a practical nonlinear optical system. The lithographically defined nano-structures offer versatile phase matching possibilities for SHG. We have presented both modal phase matching in uniform LN waveguides and quasi-phase matching in PGLN waveguides, achieving normalized conversion efficiencies of 41% $W^{-1}cm^{-2}$ and 6.8% $W^{-1}cm^{-2}$, respectively. These values are on par with PPLN in both ion-exchanged waveguides [9, 11] and recently reported SiN-LNOI hybrid waveguides [32], but are achieved with a single lithographic step, and without the need for periodic domain inversion. While phase-matched LN waveguides feature higher conversion efficiency and are easier to fabricate, PGLN waveguides provide fundamental optical modes at both input and output channels just like conventional PPLN waveguides, which is more appealing in many applications.

Further optimization on the etched sidewall roughness, including reflow of the resist, could reduce the propagation loss by at least an order of magnitude [26]. This would allow us to extend the total waveguide length beyond conventional PPLN (~ 10 cm), while the total device footprint can still be maintained within ~ $mm^2$ due to the small bending radii (< 20 μm) allowed by our etched waveguides. Reduced intrinsic propagation loss will also improve the global maximum conversion efficiencies for PGLN waveguides, as we have discussed in Section 2.2. Our simulation shows that, using a sinusoidal width modulation in PGLN waveguides could reduce the radiation loss due to leaky Bloch mode by a factor of 2 for pump wavelength and two orders of magnitude for SH wavelength. Moreover, both methods could be used in a micro-resonator geometry, which would further boost the conversion efficiency by several orders of magnitude, as is already demonstrated in other materials (e.g. AlN [19]). Our work could be a crucial step towards on-chip quantum wavelength conversion at the single-photon level. The nanofabricated LN devices are also promising for applications at short wavelengths (UV-visible) where conventional PPLN approaches become challenging since much smaller poling periods are required.

## *Acknowledgements*

This work was supported in part by National Science Foundation (NSF) (ECCS-1609549); AFOSR MURI on Quantum Memories (FA9550-12-1-0025); National Natural Science Foundation of China (11374289, 61590932); Fundamental Research Funds for the Central Universities of China; and National Nanotechnology Infrastructure Network (NNIN) Research Experience for Undergraduates (REU) program.

The authors thank Zin Lin and Dr. Mian Zhang for valuable discussions, I-Chun Huang for using his APD. We thank Dr. Hui Hu from NANOLN for helpful discussions. Device fabrication was performed at the Center for Nanoscale Systems (CNS) at Harvard University.

## *References*

1. W. J. Kozlovsky, C. D. Nabors, and R. L. Byer, "Efficient second harmonic generation of a diode-laser-pumped CW Nd:YAG laser using monolithic MgO:LiNbO$_3$ external resonant cavities," IEEE J. of Quantum Electron. **24**, 913-919 (1988).
2. K. L. Vodopyanov, M. M. Fejer, X. Yu, J. S. Harris, Y.-S. Lee, W. C. Hurlbut, V. G. Kozlov, D. Bliss, and C. Lynch, "Terahertz-wave generation in quasi-phase-matched GaAs," Appl. Phys. Lett. **89**, 141119 (2006).
3. J. P. Meyn, C. Laue, R. Knappe, R. Wallenstein, and M. M. Fejer, "Fabrication of periodically poled lithium tantalate for UV generation with diode lasers," Appl. Phys. B **73**, 111-114 (2001).
4. A. Vaziri, G. Weihs, and A. Zeilinger, "Experimental two-photon, three-dimensional entanglement for quantum communication," Phys. Rev. Lett. **89**, 240401 (2002).
5. S. Tanzilli, W. Tittel, M. Halder, O. Alibart, P. Baldi, N. Gisin, and H. Zbinden, "A photonic quantum information interface," Nature **437**, 116-120 (2005).
6. S. Zaske, A. Lenhard, C. A. Keßler, J. Kettler, C. Hepp, C. Arend, R. Albrecht, W. M. Schulz, M. Jetter, P. Michler, and C. Becher, "Visible-to-telecom quantum frequency conversion of light from a single quantum emitter," Phys. Rev. Lett. **109**, 147404 (2012).
7. M. M. Fejer and R. L. Byer, "Second-harmonic generation of green light in periodically poled planar lithium niobate waveguide," Electron. Lett. **25**, 174-175 (1989).
8. T. Sugita, K. Mizuuchi, Y. Kitaoka, and K. Yamamoto, "31%-efficient blue second-harmonic generation in a periodically poled MgO:LiNbO$_3$ waveguide by frequency doubling of an AlGaAs laser diode," Opt. Lett. **24**, 1590-1592 (1999).

9. K. R. Parameswaran, R. K. Route, J. R. Kurz, R. V. Roussev, M. M. Fejer, and M. Fujimura, "Highly efficient second-harmonic generation in buried waveguides formed by annealed and reverse proton exchange in periodically poled lithium niobate," Opt. Lett. **27**, 179-181 (2002).

10. R. V. Roussev, C. Langrock, J. R. Kurz, and M. M. Fejer, "Periodically poled lithium niobate waveguide sum-frequency generator for efficient single-photon detection at communication wavelengths," Opt. Lett. **29**, 1518-1520 (2004).

11. L. Ming, C. B. E. Gawith, K. Gallo, M. V. O'Connor, G. D. Emmerson, and P. G. R. Smith, "High conversion efficiency single-pass second harmonic generation in a zinc-diffused periodically poled lithium niobate waveguide," Opt. Express **13**, 4862-4868 (2005).

12. D. N. Nikogosyan, *Nonlinear Optical Crystals: A Complete Survey* (Springer-Science, New York, 2005).

13. Y. N. Korkishko, V. A. Fedorov, T. M. Morozova, F. Caccavale, F. Gonella, and F. Segato, "Reverse proton exchange for buried waveguides in $LiNbO_3$," J. Opt. Soc. Am. A **15**, 1838-1842 (1998).

14. P. Del'Haye, A. Schliesser, O. Arcizet, T. Wilken, R. Holzwarth, and T. J. Kippenberg, "Optical frequency comb generation from a monolithic microresonator," Nature **450**, 1214-1217 (2007).

15. A. C. Turner, M. A. Foster, A. L. Gaeta, and M. Lipson, "Ultra-low power parametric frequency conversion in a silicon microring resonator," Opt. Express **16**, 4881-4887 (2008).

16. J. S. Levy, A. Gondarenko, M. A. Foster, A. C. Turner-Foster, A. L. Gaeta, and M. Lipson, "CMOS-compatible multiple-wavelength oscillator for on-chip optical interconnects," Nat. Photon. **4**, 37-40 (2010).

17. C. Xiong, W. Pernice, K. K. Ryu, C. Schuck, K. Y. Fong, T. Palacios, and H. X. Tang, "Integrated GaN photonic circuits on silicon (100) for second harmonic generation," Opt. Express **19**, 10462-10470 (2011).

18. K. Rivoire, S. Buckley, F. Hatami, and J. Vučković, "Second harmonic generation in GaP photonic crystal waveguides," Appl. Phys. Lett. **98**, 263113 (2011).

19. W. H. P. Pernice, C. Xiong, C. Schuck, and H. X. Tang, "Second harmonic generation in phase matched aluminum nitride waveguides and micro-ring resonators," Appl. Phys. Lett. **100**, 223501 (2012).

20. S. Buckley, M. Radulaski, J. Petykiewicz, K. G. Lagoudakis, J. H. Kang, M. Brongersma, K. Biermann, and J. Vučković, "Second-harmonic generation in GaAs photonic crystal cavities in (111)B and (001) crystal orientations," ACS Photonics **1**, 516-523 (2014).

21. B. J. M. Hausmann, I. Bulu, V. Venkataraman, P. Deotare, and M. Loncar, "Diamond nonlinear photonics," Nat. Photon. **8**, 369-374 (2014).

22. S. Yamada, B. S. Song, S. Jeon, J. Upham, Y. Tanaka, T. Asano, and S. Noda, "Second-harmonic generation in a silicon-carbide-based photonic crystal nanocavity," Opt. Lett. **39**, 1768-1771 (2014).

23. G. Poberaj, H. Hu, W. Sohler, and P. Günter, "Lithium niobate on insulator (LNOI) for micro-photonic devices," Laser & Photonics Rev. **6**, 488-503 (2012).

24. C. Wang, M. J. Burek, Z. Lin, H. A. Atikian, V. Venkataraman, I. C. Huang, P. Stark, and M. Lončar, "Integrated high quality factor lithium niobate microdisk resonators," Opt. Express **22**, 30924-30933 (2014).

25. J. Lin, Y. Xu, Z. Fang, M. Wang, J. Song, N. Wang, L. Qiao, W. Fang, and Y. Cheng, "Fabrication of high-Q lithium niobate microresonators using femtosecond laser micromachining," Sci. Rep. **5**, 8072 (2015).

26. J. Wang, F. Bo, S. Wan, W. Li, F. Gao, J. Li, G. Zhang, and J. Xu, "High-Q lithium niobate microdisk resonators on a chip for efficient electro-optic modulation," Opt. Express **23**, 23072-23078 (2015).

27. A. Guarino, G. Poberaj, D. Rezzonico, R. Degl'Innocenti, and P. Gunter, "Electro-optically tunable microring resonators in lithium niobate," Nat. Photon. **1**, 407-410 (2007).

28. C. Wang, M. J. Burek, z. lin, H. A. Atikian, V. Venkataraman, I. c. Huang, P. Stark, and M. Loncar, "Integrated lithium niobate nonlinear optical devices," in *CLEO: 2015*, OSA Technical Digest (online) (Optical Society of America, 2015), FW1D.1.

29. S. Diziain, R. Geiss, M. Zilk, F. Schrempel, E.-B. Kley, A. Tünnermann, and T. Pertsch, "Second harmonic generation in free-standing lithium niobate photonic crystal L3 cavity," Appl. Phys. Lett. **103**, 051117 (2013).

30. R. Geiss, S. Saravi, A. Sergeyev, S. Diziain, F. Setzpfandt, F. Schrempel, R. Grange, E.-B. Kley, A. Tünnermann, and T. Pertsch, "Fabrication of nanoscale lithium niobate waveguides for second-harmonic generation," Opt. Lett. **40**, 2715-2718 (2015).

31. L. Cai, Y. Wang, and H. Hu, "Low-loss waveguides in a single-crystal lithium niobate thin film," Opt. Lett. **40**, 3013-3016 (2015).

32. L. Chang, Y. Li, N. Volet, L. Wang, J. Peters, and J. E. Bowers, "Thin film wavelength converters for photonic integrated circuits," Optica **3**, 531-535 (2016).

33. D. Duchesne, K. A. Rutkowska, M. Volatier, F. Légaré, S. Delprat, M. Chaker, D. Modotto, A. Locatelli, C. De Angelis, M. Sorel, D. N. Christodoulides, G. Salamo, R. Arès, V. Aimez, and R. Morandotti, "Second harmonic generation in AlGaAs photonic wires using low power continuous wave light," Opt. Express **19**, 12408-12417 (2011).

34. R. W. Boyd, *Nonlinear optics* (Academic press, 2003).

35. S. Somekh and A. Yariv, "Phase-matchable nonlinear optical interactions in periodic thin films," Appl. Phys. Lett. **21**, 140-141 (1972).

36. T. Suhara and H. Nishihara, "Theoretical analysis of waveguide second-harmonic generation phase matched with uniform and chirped gratings," IEEE J. of Quantum Electron. **26**, 1265-1276 (1990).

37. S. Saravi, S. Diziain, M. Zilk, F. Setzpfandt, and T. Pertsch, "Phase-matched second-harmonic generation in slow-light photonic crystal waveguides," Phys. Rev. A **92**, 063821 (2015).

38. R. Regener and W. Sohler, "Loss in low-finesse Ti:LiNbO$_3$ optical waveguide resonators," Appl. Phys. B **36**, 143-147 (1985).

39. M. J. Burek, C. Meuwly, R. E. Evans, M. K. Bhaskar, A. Sipahigil, S. Meesala, D. D. Sukachev, C. T. Nguyen, J. L. Pacheco, E. Bielejec, M. D. Lukin, and M. Lončar, "A fiber-coupled diamond quantum nanophotonic interface," arXiv: 1612.05285 (2016).